\documentclass[preprintnumbers,aps,prl,twocolumn,superscriptaddress,notitlepage,nofootinbib]{revtex4-1}
\usepackage{amsmath}
\usepackage{epsfig}
\usepackage{color,soul}
\usepackage{multirow}
\def\bea#1\eea{\begin{align}#1\end{align}}

\newcommand{\bef}{\begin{figure}[h!tb]\centering}
\newcommand{\eef}{\end{figure}}
\newcommand{\be}{\begin{equation}}
\newcommand{\ee}{\end{equation}}
\newcommand{\f}{\frac}

\begin{document}
\preprint{JLAB-THY-19-2896}
\title{Factorization of jet cross sections in heavy-ion collisions}
                                
\author{Jian-Wei Qiu}
\email{jqiu@jlab.org}
\affiliation{Theory Center, Jefferson Laboratory, Newport News, Virginia 23606, USA}

\author{Felix Ringer}
\email{fmringer@lbl.gov}
\affiliation{Nuclear Science Division, Lawrence Berkeley National Laboratory, Berkeley, California 94720, USA}

\author{Nobuo Sato}
\email{nsato@jlab.org}
\affiliation{Theory Center, Jefferson Laboratory, Newport News, Virginia 23606, USA}
\affiliation{Department of Physics, Old Dominion University, Norfolk, Virginia 23529, USA}
                   
\author{Pia Zurita}
\email{maria.zurita@ur.de}
\affiliation{Institut f\"ur Theoretische Physik, Universit\"at Regensburg, 93040 Regensburg, Germany}

\date{\today}         

\begin{abstract}
We propose a new phenomenological approach to establish QCD
factorization of jet cross sections in the heavy-ion environment.
Starting from a factorization formalism in proton-proton collisions,
we introduce medium modified jet functions to capture the leading
interaction of jets with the hot and dense QCD medium. A global
analysis using a Monte Carlo sampling approach is performed in order
to reliably determine the new jet functions from the nuclear
modification factor of inclusive jets at the LHC. We find that gluon
jets are significantly more suppressed due to the presence of the
medium than quark jets. In addition, we observe that the jet radius
dependence is directly related to the relative suppression of quark
and gluon jets. Our approach may help to improve the extraction of
medium properties from data.
\end{abstract}

\date{\today}

\maketitle
{\it Introduction.} In heavy-ion collisions (HIC) at the LHC and RHIC
hard probes such as highly energetic jets and hadrons are used to
extract information about the created hot and dense QCD medium, the
quark-gluon plasma (QGP)~\cite{Gyulassy:2003mc,Accardi:2004gp}.  Since
no parton is observed in isolation, QCD factorization is
necessary to separate the physics that live at different scales and to
link the quarks and gluons in hard collisions to the hadrons 
observed in the detectors~\cite{Collins:1989gx}.  The factorization
has been applied successfully at collider and fixed target
experiments. In particular, it is possible to consistently extract
universal parton distribution functions (PDFs) within global analyses
from different processes and experiments
~\cite{Harland-Lang:2017ytb,Dulat:2015mca,Alekhin:2017kpj,Ball:2017nwa,Accardi:2016qay}.
These phenomenological results support the validity of QCD
factorization in proton-proton ($p+p$) collisions and the universality
of PDFs, which ensures the predictive power of the approach. 
 
However, QCD factorization for observables in hadron-hadron collisions
is an approximation with corrections typically suppressed by inverse
powers of the large momentum transfer of the hard scattering.
Although the proof of factorization theorems at the leading power of
the large momentum transfer is independent of the details of the
identified hadrons, the corrections to the factorized formalism are
very much sensitive to what hadrons are colliding or observed in the
final-state.  This is because the subleading power contributions to
the hadronic observables are very sensitive to QCD multiple
scattering and, therefore, depend on where the collision is taking
place, in a proton, a heavy ion, or a QGP-like hot medium.  That is,
the kinematic regime where the leading power  formalism is applicable
could be very different for $p+p$, proton-ion, or ion-ion ($A+A$)
collisions.  Tremendous efforts have been devoted to study multiple
scatterings in QCD, and their medium modifications to hadronic
observables, such as jet quenching, from which medium properties were
extracted~\cite{Gyulassy:1993hr,Baier:1996sk,Zakharov:1996fv,Gyulassy:2000er,Wang:2001ifa,Arnold:2002ja,Qiu:2004da,Armesto:2011ht,Burke:2013yra,Andres:2019eus}. Since only the first subleading power contributions to hadronic
observables can be factorized to all orders in perturbative QCD (pQCD)
in a similar way to the leading power
contributions~\cite{Qiu:1990xy,Botts:1990uy,Qiu:2003cg}, some kind of
model dependence is needed for studying QCD multiple scatterings which
can introduce a model bias of the extracted medium properties.  

Given the importance of jet quenching observables for extracting QGP
properties in HIC, we explore in this Letter the validity   of the
leading power, model independent QCD factorization formalism for
inclusive single jet production in $A+A\to {\rm jet}+X$.  Using the
leading power factorization formalism and the same partonic hard parts
and jet evolution for $p+p$ collisions, we demonstrate for the first
time that we are able to interpret the jet suppression $R_{\rm
AA}^{\rm jet}$ data from the LHC by fitting medium induced jet
functions. We use a Monte Carlo (MC) sampling approach to reliably
determine the new medium modified jet functions and to identify the
kinematic regime where the factorization approach is indeed feasible.
This data driven approach to verify factorization in HIC may open a
new door toward extractions of medium properties with a reduced model
bias. Eventually, a global analysis of different observables is needed
to establish more rigorously the universality of these nonperturbative
functions; and a consistent treatment of medium sensitive power
corrections is required to extend the predictive power of our
formalism to HIC at lower energies.

{\it Theoretical framework.}  Inclusive single jet cross section in $p+p$
collisions, differential in the transverse momentum $p_T$ and rapidity $\eta$,
can be factorized as~\cite{Ellis:1993tq}
\begin{eqnarray}
\label{eq:jet}
\f{d\sigma^{pp\to{\rm jet}+X}}{dp_T d\eta}
&=&\sum_{ab}f_{a/p} \otimes f_{b/p} \otimes {\cal H}_{ab}^{\rm jet}
\\
&\ & {\hskip -0.6in}
= \sum_{ab}f_{a/p} \otimes f_{b/p} \otimes \Bigl[
\sum_c \hat{\sigma}_{ab\to c} \otimes J_{c} 
+\hat{\sigma}_{ab}^{\rm Jet}
\Bigr] \,.
\label{eq:factorization}
\end{eqnarray}
Here $f_{i/p}(x_i)$ with $i=a,b$ are the PDFs, $\otimes$ indicates
appropriate integrals over parton momentum fractions and ${\cal
H}_{ab}^{\rm jet}$ are partonic hard parts for the colliding partons
of flavor $a$ and $b$ to produce the observed jet, which are
perturbatively calculable depending on the jet algorithm.  When the
observed jet is very energetic and narrow in cone size $R$, the
partonic hard parts ${\cal H}_{ab}^{\rm jet}$ are dominated by large
logarithms in $\ln(R)$.  Since the $\ln(R)$ are due to the sensitivity
to collinear final-state radiation that forms the jet, the resummation
of $\alpha_s^n\ln^n(R)$ is needed which can be consistently achieved
by reorganizing ${\cal H}_{ab}^{\rm jet}$ analogous
to~\cite{Berger:2001wr}.  The separation of ${\cal H}_{ab}^{\rm jet}$
into a ``jet-independent" partonic hard part, $\hat{\sigma}_{ab\to
c}(z,\mu)$, for producing a parton $c$ of transverse momentum
$p^c_T=p_T/z$ at a factorization scale $\mu\sim p_T$ and a
``jet-dependent'' jet function, $J_{c}(z,p_TR,\mu)$, which accounts
for the formation of the observed jet from the parton $c$, as
indicated in Eq.~(\ref{eq:factorization}) allows for the resummation
of $\ln(R)$ terms to all
orders~\cite{Dasgupta:2014yra,Kaufmann:2015hma,Kang:2016mcy,Dai:2016hzf}.
The $\hat{\sigma}_{ab}^{\rm Jet}$ in Eq.~(\ref{eq:factorization}) are
either $R$-independent or suppressed by powers of
$R^2$~\cite{Ellis:1992qq}, and can be neglected if $R$ is sufficiently
small. Therefore, we do not consider $\hat{\sigma}_{ab}^{\rm Jet}$ in
our analysis.  Terms which are further suppressed by inverse powers of
$p_T$ are also neglected as they are beyond the factorization formulas
in Eqs.~(\ref{eq:jet}) and (\ref{eq:factorization}).  

When ${\cal H}_{ab}^{\rm jet}$ is reorganized for deriving
Eq.~(\ref{eq:factorization}), we can choose the ``jet-independent''
$\hat{\sigma}_{ab\to c}(z,\mu)$ to be the same as the partonic hard
part for inclusive single hadron production at high
$p_T$~\cite{Aversa:1988vb,Jager:2002xm}, which is factorized
as~\cite{Nayak:2005rt},
\begin{equation}
\label{eq:hadron}
\f{d\sigma^{pp\to h+X}}{dp_T d\eta}
=\sum_{abc}f_{a/p} \otimes f_{b/p} \otimes \hat{\sigma}_{ab\to c}(z,\mu) \otimes D^h_{c}(z,\mu) \, .
\end{equation}
Here $D_c^h$ are the single hadron fragmentation functions (FFs), and
the dependence on the initial-state partonic momentum fractions and
the factorization scale are left implicit.  Since the physically
observed cross section on the left hand side is independent of the
factorization scale, the $\mu$-dependence of the FFs follows the DGLAP
evolution where the evolution kernels are uniquely determined by the
$\mu$-dependence of  $\hat{\sigma}_{ab\to c}(z,\mu)$, order-by-order
in pQCD.  Since $\hat{\sigma}_{ab\to c}(z,\mu)$ is the same in both
Eqs.~(\ref{eq:factorization}) and (\ref{eq:hadron}), the jet functions
obey the same DGLAP evolution equation,
\be\label{eq:dglap}
\mu\f{d}{d\mu}J_c(z,p_T R,\mu)=\sum_d P_{dc}(z) \otimes J_d(z,p_T R,\mu)\, ,
\ee
%
with the same  $P_{dc}(z)$ as for FFs. 
Solving the DGLAP evolution equation from the jet invariant mass
$\mu_J\sim p_T R$ to $\mu\sim p_T$, the scale of the hard collision,
effectively resums single logarithms in the jet radius
$\alpha_s^n\ln^n(R)$. 
Although the $J_c$ in Eq.~(\ref{eq:factorization}) play the same role
as the $D_c^h$ in Eq.~(\ref{eq:hadron}), they  are calculable
order-by-order in pQCD, while the FFs are nonperturbative and need to
be extracted from experimental data.  The factorized formalism in
Eq.~(\ref{eq:factorization}) has been successfully tested for single
inclusive jet production in $p+p$ collisions at the
LHC~\cite{Liu:2018ktv}.

When we apply Eq.~(\ref{eq:factorization}) to narrow-cone jet
production in HIC, only the PDFs and the jet functions should be
modified since $\hat{\sigma}_{ab\to c}$ is insensitive to the
long-distance physics.  Although nuclear PDFs (nPDFs) differ from
nucleon PDFs, we note that their impact is generally small which is
consistent with the expectation that jet quenching is a final state
effect~\cite{Adams:2003im,Adler:2003ii,Khachatryan:2016xdg,Acharya:2017okq}.
That is, the main source of jet quenching is likely to be multiple
scattering and medium induced energy loss as the jet traverses the
QGP, which modify the $J_{c}$ in $p+p$ collisions into medium
sensitive and nonperturbative jet functions ($J_c^{\rm med}$), 
\be
J_c(z,p_T R,\mu) \to J_c^{\rm med}(z,p_T R,\mu) \,.
\ee
The factorization of jet production in HIC in terms of $J_c^{\rm med}$
was first proposed in~\cite{Kang:2017frl,Li:2018xuv} where a model
calculation~\cite{Ovanesyan:2011xy} was performed.
In~\cite{He:2018gks}, the medium modification was taken to be a
function of the jet $p_T$ and the jet energy loss was determined at
the cross section level. Other recent data driven approaches can be
found
in~\cite{Casalderrey-Solana:2018wrw,Ke:2018tsh,Sirimanna:2019bgl}.
The factorization formalism in Eq.~(\ref{eq:factorization}) with
$J_c^{\rm med}$ allows us to directly work at the parton level to
study how the parton shower (PS) gets modified due to the presence of
the QGP. In~\cite{Brewer:2018dfs} a new approach at the level of jet
cross sections was introduced.

\begin{figure}[t]
\includegraphics[width=8.7cm]{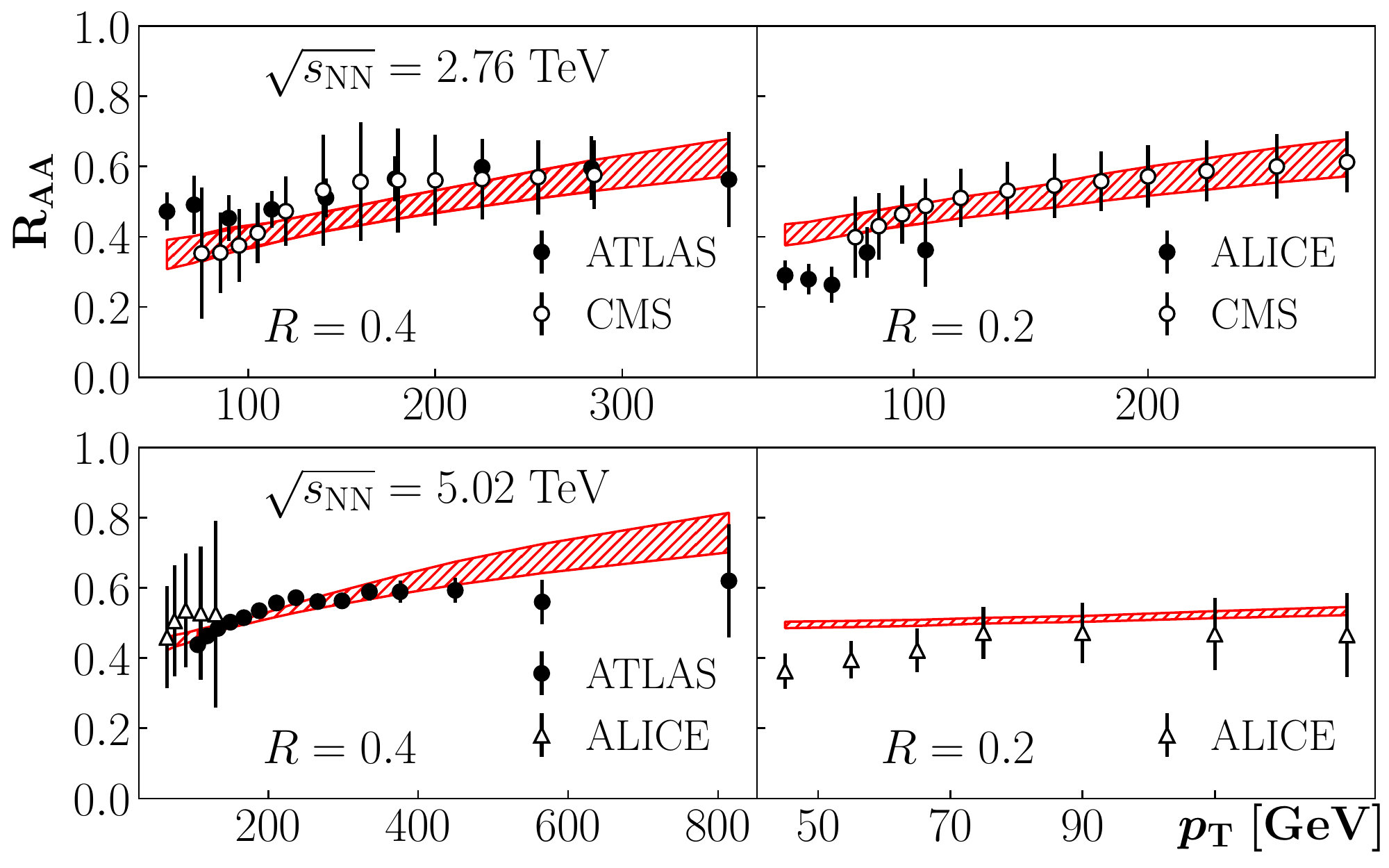}
\caption{
  The $R_{\rm AA}^{\rm jet}$ for inclusive jet 
  production (0-10$\%$ centrality) at $\sqrt{s_{\rm NN}}=2.76$~TeV
  (upper panels) and $\sqrt{s_{\rm NN}}=5.02$~TeV (lower panels).
  We show the comparison with the  data from
  ALICE~\cite{Adam:2015ewa,Mulligan:2018nnq},
  ATLAS~\cite{Aad:2012vca,Aaboud:2018twu} and
  CMS~\cite{Khachatryan:2016jfl}.\label{fig:RAA}}
\vspace*{-0.3cm}
\end{figure}

We stress that the proposed factorization approach is complementary to
others in the literature, see for example~\cite{Wang:2016opj} and
references therein.  In-medium calculations based on analytical
techniques or PS event generators rely on some kind of factorization
in HIC.  With the leading power factorization formalism used here, our
approach reduces the model bias to a minimum.  The success of our
framework, as demonstrated below, can help us to focus on how the
medium modifies the jet functions in order to develop microscopic
models of the QGP and its interaction with hard probes.

To be consistent with QCD factorization at leading power, we leave the
DGLAP evolution equation and the corresponding kernels in
Eq.~(\ref{eq:dglap}) unmodified and only change the initial condition
of the evolution.  In a PS picture this corresponds to keeping the
shower between the hard scale $p_T$ and the jet scale $p_T R$ to be
the same as that in the vacuum. Instead, only the physics at lower
scales is affected by the QCD medium, which is captured effectively by
fitting $J_c^{\rm med}$ to the data at the jet scale $\mu_J\sim p_T
R$.  This is consistent for example with the PS developed
in~\cite{Zapp:2008gi,Schenke:2009gb,Wang:2013cia,Cao:2017qpx,Cao:2017zih}
where the shower is unmodified relative to the vacuum case at
sufficiently large scales. In principle, it is possible to extend our
calculation to include a medium modified evolution which can be
constrained from data and which we leave for future
work~\cite{Armesto:2009fj}. 

Our analysis here is similar to the global analyses of
nPDFs~\cite{Eskola:2016oht,deFlorian:2011fp,Kovarik:2015cma} and
nuclear fragmentation functions in cold nuclear
matter~\cite{Sassot:2009sh}. Since the $J_{c}$  are perturbatively
calculable, we choose an ansatz where the $J_{c}^{\rm med}$ are
written in terms of the vacuum ones convolved with weight functions
$W_c(z)$,
\be\label{eq:mediumjet}
J_c^{\rm med}(z,p_T R,\mu_J) = W_c(z)  \otimes J_c(z,p_T R,\mu_J) \,.
\ee
This approach effectively assumes that the QGP introduces a
factorizable modification of the $J_{c}$, which recovers the vacuum
case, for example, for very peripheral interactions, by having
$W_c(z)\to\delta(1-z)$. We adopt the following flexible
parametrization,
\be\label{eq:weightfct}
W_c(z)=\epsilon_c\delta(1-z)+N_c \, z^{\alpha_c} (1-z)^{\beta_c}\,,
\ee
for the weight functions. As the dependence on the factorization scale
$\mu$ of the $J_{c}$ is associated with the leading $\ln(R)$
contribution to the jet cross sections, one finds 
$\mu \f{d}{d\mu} \int_0^1dz\, z\, J_c(z,p_T R,\mu) 
\propto \sum_d \int_0^1dz\, z\,P_{dc}(z) = 0$.  
That is, the first moment of $J_{c}$ is independent
of the factorization scale. Due to momentum conservation of the
fragmenting parton $p_T^c$, the $J_{c}$ satisfy the sum rule
\be\label{eq:normalization}
\int_0^1dz\, z\, J_c(z,p^c_T R,\mu) = 1
\, ,
\ee
which provides constraints for the evolution of the jet functions both
in the vacuum and the medium. The convolution structures in
Eqs.~(\ref{eq:factorization}) and~(\ref{eq:mediumjet}) can be handled
conveniently in Mellin moment space~\cite{deFlorian:2009vb}. The
parameters of the weight functions are determined by a MC sampling of
the likelihood function 
$
{\cal \rho}(\bf{a}|{\rm data}) \propto
{\cal L}(\bf{a},{\rm data})\pi({\bf{a}})
$
with 
$
{\cal L}(\bf{a},{\rm data})
  =\exp\left[-\frac{1}{2}\chi^2(\bf{a},{\rm data})
       \right]
$,
where the data resampling method (NNPDF~\cite{Ball:2017nwa},
JAM~\cite{Accardi:2016qay}) is used in order to obtain the MC ensemble
for the parameters. 

{\it Phenomenological results.} We consider inclusive jet data in HIC
from the LHC, with the nuclear modification factor defined as
\be
R_{\rm AA}^{\rm jet}=
 \f{d\sigma^{\text{PbPb}\to\text{jet}+X}}
   {\langle T_{\rm AA}\rangle \, d\sigma^{pp\to\text{jet}+X}} \,,
\ee
where $\langle T_{\rm AA}\rangle$ is the average nuclear overlap
function over a given $A+A$ centrality class~\cite{dEnterria:2003xac}.
The $J_{c}^{\rm med}$ need to be extracted separately for different
centrality classes and center-of-mass (CM) energies. We include all
available data sets from the LHC and limit ourselves here to the most
central collisions (0-10$\%$). At $\sqrt{s_{\rm NN}}=2.76$~TeV we
include the data from ALICE~\cite{Adam:2015ewa},
ATLAS~\cite{Aad:2012vca} and CMS~\cite{Khachatryan:2016jfl} and at
$\sqrt{s_{\rm NN}}=5.02$~TeV we consider the ATLAS data
of~\cite{Aaboud:2018twu} and the preliminary ALICE data
of~\cite{Mulligan:2018nnq}. For all data sets the anti-k$_T$
algorithm~\cite{Cacciari:2008gp} was used with jet radii in the range
of $R=0.2$-$0.4$. The data sets cover different rapidity ranges which
we take into account without listing them here.  We add correlated and
uncorrelated uncertainties in quadrature. For all numerical results
presented here we use the CT14 PDF set of~\cite{Dulat:2015mca}, and we
work at next-to-leading order supplemented with resummation at
next-to-leading logarithmic accuracy.
\begin{figure}[t]
\includegraphics[width=8.7cm]{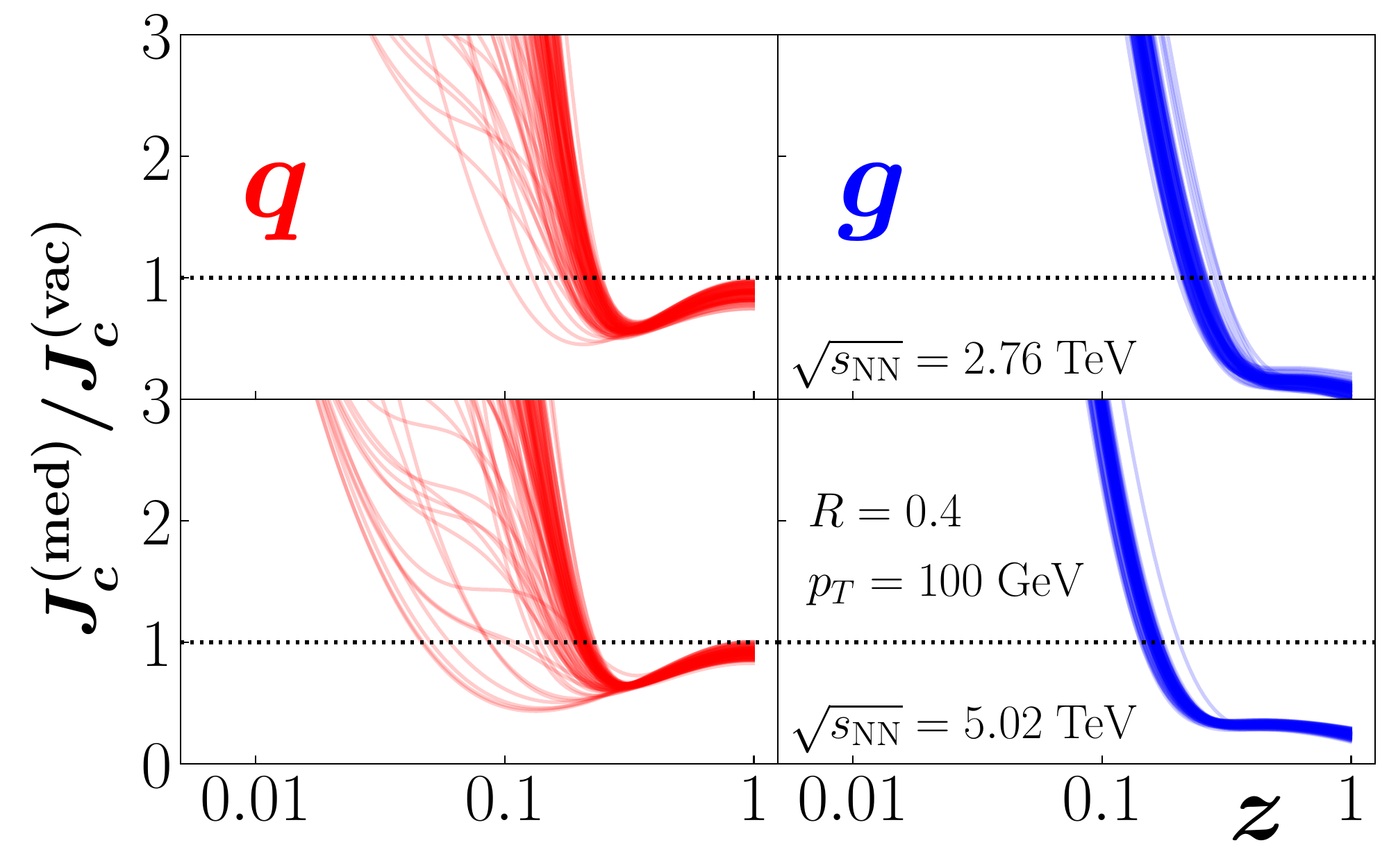}
\vspace*{-.6cm}
\caption{
  Ratio of the extracted  $J_{c}^{\rm med}$ and $J_{c}$ at
  $\sqrt{s_{\rm NN}}=2.76$~TeV (upper panels) and $\sqrt{s_{\rm NN}}=5.02$~TeV 
  (lower panels) evaluated for $R=0.4$ jets at $\mu=p_T=100$~GeV for
  quarks (left) and gluons (right). \label{fig:jetfcts}
}
\vspace*{-.3cm}
\end{figure}
In Fig.~\ref{fig:RAA}, we present a comparison of data from the LHC
for the $R_{\rm AA}^{\rm jet}$ and our theoretical results using the
fitted $J_{c}^{\rm med}$. We show the results at~$\sqrt{s_{\rm
NN}}=2.76$~TeV (upper panels) and $\sqrt{s_{\rm NN}}=5.02$~TeV (lower
panels). For both CM energies we find good agreement with a
$\chi^2/$d.o.f. of 1.1 (2.76~TeV) and 1.7 (5.02~TeV). At low jet $p_T$
there may be an indication for a medium modified DGLAP evolution,
while the precision of current data does not require it yet.  More
insights could be obtained from analyzing hadron and jet substructure
observables. 

In Fig.~\ref{fig:jetfcts}, we present the ratio of the extracted
$J_{c}^{\rm med}$ and their vacuum analogues for $\sqrt{s_{\rm
NN}}=2.76$~TeV (upper panels) and $\sqrt{s_{\rm NN}}=5.02$~TeV (lower
panels) separately for quark (left) and gluon (right) jets with
$R=0.4$ at the scale $\mu=p_T=100$~GeV. We find that the uncertainty
at the higher CM energy is reduced significantly. This is mainly due
to the very precise data set from ATLAS at
5.02~TeV~\cite{Aaboud:2018twu} which dominates the corresponding fit. 

At large-$z$ the suppression of the jet functions indicates that it is
less likely to form a jet carrying a large momentum fraction of the
fragmenting parton in HIC. This is consistent with existing parton
energy loss models~\cite{Baier:1996sk,Gyulassy:2000er}.  The
suppression of $J_{c}^{\rm med}$ at large-$z$ leads to the suppression
of the inclusive jet cross section. On the other hand, the large-$z$
suppression is compensated by an enhancement at small-$z$, see also
Eq.~(\ref{eq:normalization}).  We note that the HIC jet data puts more
significant constraints on the large-$z$ region of the $J_{c}^{\rm
med}$ . This is due to the convolution structure of the jet cross
section, which forces the phase space with a combination of small
$x_{a,b}$ and large $z$ to dominate the jet production rate.  A
possibility to constrain the small-$z$ behavior more directly is the
measurement of the energy distribution of inclusive
subjets~\cite{Kang:2017mda}.  
\begin{figure}[t]
\vspace*{-.2cm}
\includegraphics[width=8.7cm]{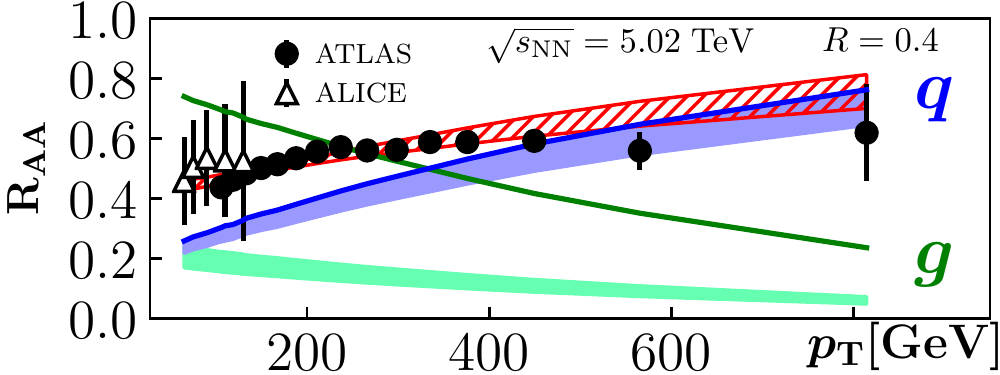}
\vspace*{-.2cm}
\caption{
  The suppression of the quark (blue) and gluon (green) cross sections
  for the lower left panel of Fig.~\ref{fig:RAA} as an example. The
  individual suppression (bands) can be seen relative to the vacuum
  fractions (lines). \label{fig:RAAqg}}
\vspace*{-.3cm}
\end{figure} 

In Fig.~\ref{fig:jetfcts} we also observe a significant difference
between 
$J_{q}^{\rm med}$ and $J_{g}^{\rm med}$ where gluon
jets are significantly more suppressed at large-$z$ than quark jets.
This behavior is generally expected from model calculations. In fact,
we find that it is not possible to fit the experimental data with the
same weight function for quarks and gluons in
Eq.~(\ref{eq:weightfct}), while retaining a probabilistic
interpretation (positivity) of the $J_{c}^{\rm med}$.  We investigated
this large difference at the level of the cross section which requires
us to define quark and gluon jets beyond leading-order.  This can be
achieved by introducing the jet functions $J_{cd}$ that not only keep
track of the parton $c$ initiating the jet but also of the flavor
content $d=q,g$ such that~\cite{Banfi:2006hf,Cal:2019hjc}
\be\label{eq:qgfractions}
\sum_d J_{cd}(z,p_T R,\mu)=J_c(z,p_T R,\mu) \,.
\ee
In Fig.~\ref{fig:RAAqg} we show the separation of the vacuum cross
section into quark (blue line) and gluon (green line) jets using the
$\sqrt{s_{\rm NN}}=5.02$~TeV setup (lower left panel of
Fig.~\ref{fig:RAA}) along with the corresponding separation in the
medium (blue and green bands). We observe that gluon jets are
significantly more suppressed than quark jets in the medium. Some jet
substructure observables indeed support this observation, see for
example~\cite{Acharya:2018uvf,Aaboud:2018hpb,Sirunyan:2018qec,Spousta:2015fca}.
In the future it will be possible to better pin down differences
between quark and gluon jets by including $\gamma/Z+{\rm jet}$
~\cite{Sirunyan:2017qhf,Aaboud:2018anc} and 
${\rm hadron}+{\rm jet}$~\cite{Adam:2015doa,Adamczyk:2017yhe}  
data in a global analysis.  We thus conclude that the leading power
factorization formalism with medium jet functions not only captures
the feature of in-medium interactions of jets with the QGP but also
allows for a clear physical interpretation.
\begin{figure}[t]
\includegraphics[width=8.7cm]{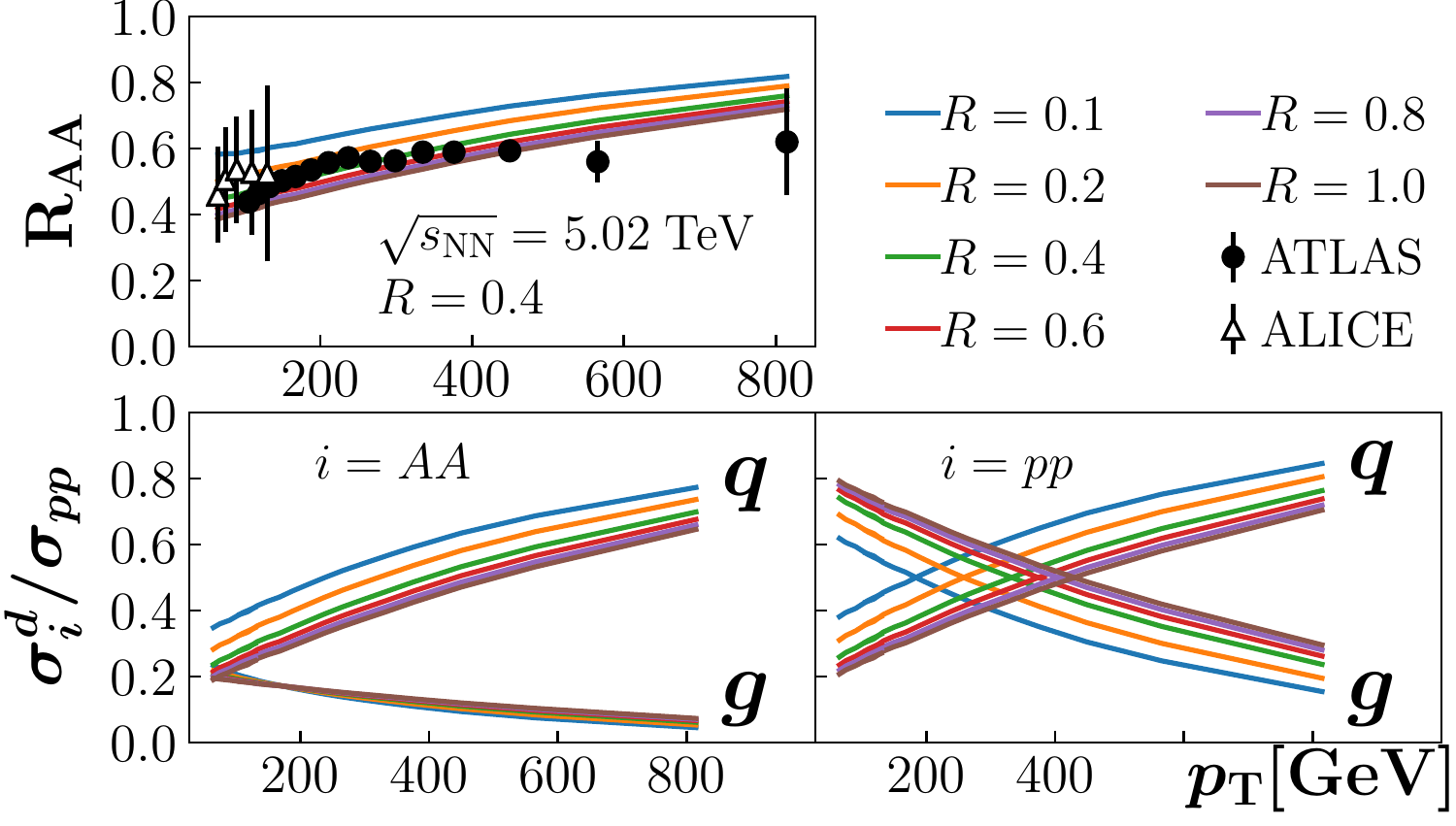}
\vspace*{-.5cm}
\caption{
  The dependence of the $R_{\rm AA}^{\rm jet}$ at $\sqrt{s_{\rm
  NN}}=5.02$~TeV on the jet radius $R$ (upper panel), and quark and
  gluon jet contributions $\sigma_i^d/\sigma_{pp}$ with $d=q,g$, see Eq.~(\ref{eq:qgfractions}), in the medium $i=AA$ (lower left) and vacuum $i=pp$ (lower
  right).\label{fig:R_dependence}}
\vspace*{-.2cm}
\end{figure}

An intriguing aspect of jet quenching studies is the jet radius
dependence. While the current experimental data remains inconclusive,
different model calculations in the literature predict the $R_{\rm
AA}^{\rm jet}$ to either increase or decrease with $R$. In general, a
non-monotonic behavior is expected:~the $R_{\rm AA}^{\rm jet}$
increases at both formal limits $R\to 0,\infty$.  In the limit $R\to
0$, the $R_{\rm AA}^{\rm jet}$ is expected to approach the hadron
$R_{\rm AA}^h$ which is generally above the $R_{\rm AA}^{\rm
jet}$~\cite{Khachatryan:2016odn}.  For large $R$ the energy lost by
partons due to medium interactions should eventually all be contained
in a very large cone. However, both limits are formally not covered by
the factorization formalism in Eq.~(\ref{eq:factorization}).  For
$R\to 0$, the jet scale $\mu_J\sim p_T R \to 0$, and the evolution
starts at $\mu_J\sim 1$~GeV with a nonperturbative $J_{c}$.  For the
experimentally accessible $R$ values it is thus a priori not clear if
the $R_{\rm AA}^{\rm jet}$ increases or decreases with $R$. In
Fig.~\ref{fig:R_dependence} we show the $R$-dependence obtained within
our framework at $\sqrt{s_{\rm NN}}=5.02$~TeV. In the vacuum the gluon
fraction of the jets decreases with smaller $R$, caused by more phase
space to evolve and the $J_{g}$ evolving faster,  which leads to the
increase of the quark fraction (lower right).  In the medium, gluon
jets are more significantly quenched (lower left), which is why the
$R_{\rm AA}^{\rm jet}$ (upper panel) effectively inherits the
$R$-dependence of the quark jets.  It will be interesting to see if
these findings will be confirmed by more precise data in the future. 

{\it Conclusions.} In this Letter, we proposed an approach to
phenomenologically establish QCD factorization of jet cross sections
in HIC.  We considered inclusive jet production at the LHC and found
that it is indeed possible to describe the $R_{\rm AA}^{\rm jet}$ by
the leading power factorization formalism for $p+p$ collisions with
medium modified jet functions.  Our results thus support the notion of
QCD factorization in the HIC environment. Since our framework operates
at the parton level, it is possible to separate quark and gluon jets.
We found that gluon jets are significantly more suppressed than quark
jets; and there is a direct link between the relative suppression of
quark and gluon jets and the jet radius dependence of the $R_{\rm
AA}^{\rm jet}$.

In the future it will be important to investigate universality aspects
of the jet functions by analyzing $\gamma/Z$ tagged jet data as well
as hadron and jet substructure observables in a similar way. The
intuitive physical interpretation of the extracted medium jet
functions may facilitate comparisons with model calculations available
in the literature.  Our proposed factorization approach helps to
identify the impact of the medium modification at the parton level,
and   may serve as guidance for constructing microscopic models of the
QGP and its interaction with hard probes. We hope that the
factorization framework may help to explore how the formation and the
evolution of a parton shower gets modified due to the presence of the
hot and dense QCD medium created in HIC, from which the properties of
the QGP can be better extracted.

{\it Acknowledgement.} We thank Y.-T.~Chien, R.~Elayavalli,
Z.-B.~Kang, K.~Lee, Y.-J.~Lee, A.~Majumder, Y.~Mehtar-Tani,
M.~Ploskon, A.~Sickles, M.~Spousta, I.~Vitev, W.~Vogelsang,
W.~Waalewijn and F.~Yuan for very helpful discussions, and
C.~Andres, F.~Dominguez, P.~Jacobs, J.~Mulligan and X.-N.~Wang also for valuable
suggestions to the manuscript. This work was supported in part by U.S.
Department of Energy under Contract Nos.\ DE-AC05-06OR23177,
DE-AC0205CH11231 and DE-SC0012704, and the LDRD program at LBNL.

\bibliographystyle{h-physrev}
\bibliography{bibliography}

\begin{thebibliography}{10}

\bibitem{Gyulassy:2003mc}
M.~Gyulassy, I.~Vitev, X.-N. Wang, and B.-W. Zhang,
\newblock p. 123 (2003), nucl-th/0302077.

\bibitem{Accardi:2004gp}
A.~Accardi {\em et~al.},
\newblock (2004), hep-ph/0310274.

\bibitem{Collins:1989gx}
J.~C. Collins, D.~E. Soper, and G.~F. Sterman,
\newblock Adv. Ser. Direct. High Energy Phys. {\bf 5}, 1 (1989),
  hep-ph/0409313.

\bibitem{Harland-Lang:2017ytb}
L.~A. Harland-Lang, A.~D. Martin, and R.~S. Thorne,
\newblock Eur. Phys. J. {\bf C78}, 248 (2018), 1711.05757.

\bibitem{Dulat:2015mca}
S.~Dulat {\em et~al.},
\newblock Phys. Rev. {\bf D93}, 033006 (2016), 1506.07443.

\bibitem{Alekhin:2017kpj}
S.~Alekhin, J.~Bl{\"u}mlein, S.~Moch, and R.~Placakyte,
\newblock Phys. Rev. {\bf D96}, 014011 (2017), 1701.05838.

\bibitem{Ball:2017nwa}
NNPDF, R.~D. Ball {\em et~al.},
\newblock Eur. Phys. J. {\bf C77}, 663 (2017), 1706.00428.

\bibitem{Accardi:2016qay}
A.~Accardi, L.~T. Brady, W.~Melnitchouk, J.~F. Owens, and N.~Sato,
\newblock Phys. Rev. {\bf D93}, 114017 (2016), 1602.03154.

\bibitem{Gyulassy:1993hr}
M.~Gyulassy and X.-n. Wang,
\newblock Nucl. Phys. {\bf B420}, 583 (1994), nucl-th/9306003.

\bibitem{Baier:1996sk}
R.~Baier, Y.~L. Dokshitzer, A.~H. Mueller, S.~Peigne, and D.~Schiff,
\newblock Nucl. Phys. {\bf B484}, 265 (1997), hep-ph/9608322.

\bibitem{Zakharov:1996fv}
B.~G. Zakharov,
\newblock JETP Lett. {\bf 63}, 952 (1996), hep-ph/9607440.

\bibitem{Gyulassy:2000er}
M.~Gyulassy, P.~Levai, and I.~Vitev,
\newblock Nucl. Phys. {\bf B594}, 371 (2001), nucl-th/0006010.

\bibitem{Wang:2001ifa}
X.-N. Wang and X.-f. Guo,
\newblock Nucl. Phys. {\bf A696}, 788 (2001), hep-ph/0102230.

\bibitem{Arnold:2002ja}
P.~B. Arnold, G.~D. Moore, and L.~G. Yaffe,
\newblock JHEP {\bf 06}, 030 (2002), hep-ph/0204343.

\bibitem{Qiu:2004da}
J.-W. Qiu and I.~Vitev,
\newblock Phys. Lett. {\bf B632}, 507 (2006), hep-ph/0405068.

\bibitem{Armesto:2011ht}
N.~Armesto {\em et~al.},
\newblock Phys. Rev. {\bf C86}, 064904 (2012), 1106.1106.

\bibitem{Burke:2013yra}
JET, K.~M. Burke {\em et~al.},
\newblock Phys. Rev. {\bf C90}, 014909 (2014), 1312.5003.

\bibitem{Andres:2019eus}
C.~Andres, N.~Armesto, H.~Niemi, R.~Paatelainen, and C.~A. Salgado,
\newblock (2019), 1902.03231.

\bibitem{Qiu:1990xy}
J.-W. Qiu and G.~F. Sterman,
\newblock Nucl. Phys. {\bf B353}, 137 (1991).

\bibitem{Botts:1990uy}
J.~Botts, J.-W. Qiu, and G.~F. Sterman,
\newblock Nucl. Phys. {\bf A527}, 577 (1991).

\bibitem{Qiu:2003cg}
J.-W. Qiu,
\newblock (2003), hep-ph/0305161.

\bibitem{Ellis:1993tq}
S.~D. Ellis and D.~E. Soper,
\newblock Phys. Rev. {\bf D48}, 3160 (1993), hep-ph/9305266.

\bibitem{Berger:2001wr}
E.~L. Berger, J.-W. Qiu, and X.-f. Zhang,
\newblock Phys. Rev. {\bf D65}, 034006 (2002), hep-ph/0107309.

\bibitem{Dasgupta:2014yra}
M.~Dasgupta, F.~Dreyer, G.~P. Salam, and G.~Soyez,
\newblock JHEP {\bf 04}, 039 (2015), 1411.5182.

\bibitem{Kaufmann:2015hma}
T.~Kaufmann, A.~Mukherjee, and W.~Vogelsang,
\newblock Phys. Rev. {\bf D92}, 054015 (2015), 1506.01415.

\bibitem{Kang:2016mcy}
Z.-B. Kang, F.~Ringer, and I.~Vitev,
\newblock JHEP {\bf 10}, 125 (2016), 1606.06732.

\bibitem{Dai:2016hzf}
L.~Dai, C.~Kim, and A.~K. Leibovich,
\newblock Phys. Rev. {\bf D94}, 114023 (2016), 1606.07411.

\bibitem{Ellis:1992qq}
S.~D. Ellis, Z.~Kunszt, and D.~E. Soper,
\newblock Phys. Rev. Lett. {\bf 69}, 3615 (1992), hep-ph/9208249.

\bibitem{Aversa:1988vb}
F.~Aversa, P.~Chiappetta, M.~Greco, and J.~P. Guillet,
\newblock Nucl. Phys. {\bf B327}, 105 (1989).

\bibitem{Jager:2002xm}
B.~Jager, A.~Schafer, M.~Stratmann, and W.~Vogelsang,
\newblock Phys. Rev. {\bf D67}, 054005 (2003), hep-ph/0211007.

\bibitem{Nayak:2005rt}
G.~C. Nayak, J.-W. Qiu, and G.~F. Sterman,
\newblock Phys. Rev. {\bf D72}, 114012 (2005), hep-ph/0509021.

\bibitem{Liu:2018ktv}
X.~Liu, S.-O. Moch, and F.~Ringer,
\newblock Phys. Rev. {\bf D97}, 056026 (2018), 1801.07284.

\bibitem{Adams:2003im}
STAR, J.~Adams {\em et~al.},
\newblock Phys. Rev. Lett. {\bf 91}, 072304 (2003), nucl-ex/0306024.

\bibitem{Adler:2003ii}
PHENIX, S.~S. Adler {\em et~al.},
\newblock Phys. Rev. Lett. {\bf 91}, 072303 (2003), nucl-ex/0306021.

\bibitem{Khachatryan:2016xdg}
CMS, V.~Khachatryan {\em et~al.},
\newblock Eur. Phys. J. {\bf C76}, 372 (2016), 1601.02001.

\bibitem{Acharya:2017okq}
ALICE, S.~Acharya {\em et~al.},
\newblock Phys. Lett. {\bf B783}, 95 (2018), 1712.05603.

\bibitem{Kang:2017frl}
Z.-B. Kang, F.~Ringer, and I.~Vitev,
\newblock Phys. Lett. {\bf B769}, 242 (2017), 1701.05839.

\bibitem{Li:2018xuv}
H.~T. Li and I.~Vitev,
\newblock (2018), 1811.07905.

\bibitem{Ovanesyan:2011xy}
G.~Ovanesyan and I.~Vitev,
\newblock JHEP {\bf 06}, 080 (2011), 1103.1074.

\bibitem{He:2018gks}
Y.~He, L.-G. Pang, and X.-N. Wang,
\newblock (2018), 1808.05310.

\bibitem{Casalderrey-Solana:2018wrw}
J.~Casalderrey-Solana, Z.~Hulcher, G.~Milhano, D.~Pablos, and K.~Rajagopal,
\newblock (2018), 1808.07386.

\bibitem{Ke:2018tsh}
W.~Ke, Y.~Xu, and S.~A. Bass,
\newblock Phys. Rev. {\bf C98}, 064901 (2018), 1806.08848.

\bibitem{Sirimanna:2019bgl}
C.~Sirimanna, S.~Cao, and A.~Majumder,
\newblock 2019, 1901.03635.

\bibitem{Brewer:2018dfs}
J.~Brewer, J.~G. Milhano, and J.~Thaler,
\newblock (2018), 1812.05111.

\bibitem{Adam:2015ewa}
ALICE, J.~Adam {\em et~al.},
\newblock Phys. Lett. {\bf B746}, 1 (2015), 1502.01689.

\bibitem{Mulligan:2018nnq}
ALICE, J.~Mulligan,
\newblock {Inclusive jet measurements in pp and Pb-Pb collisions with ALICE},
\newblock 2018, 1812.07681.

\bibitem{Aad:2012vca}
ATLAS, G.~Aad {\em et~al.},
\newblock Phys. Lett. {\bf B719}, 220 (2013), 1208.1967.

\bibitem{Aaboud:2018twu}
ATLAS, M.~Aaboud {\em et~al.},
\newblock Phys. Lett. {\bf B790}, 108 (2019), 1805.05635.

\bibitem{Khachatryan:2016jfl}
CMS, V.~Khachatryan {\em et~al.},
\newblock Phys. Rev. {\bf C96}, 015202 (2017), 1609.05383.

\bibitem{Wang:2016opj}
X.-N. Wang, editor,
\newblock {\em {Quark-Gluon Plasma 5}} (World Scientific, New Jersey, 2016).

\bibitem{Zapp:2008gi}
K.~Zapp, G.~Ingelman, J.~Rathsman, J.~Stachel, and U.~A. Wiedemann,
\newblock Eur. Phys. J. {\bf C60}, 617 (2009), 0804.3568.

\bibitem{Schenke:2009gb}
B.~Schenke, C.~Gale, and S.~Jeon,
\newblock Phys. Rev. {\bf C80}, 054913 (2009), 0909.2037.

\bibitem{Wang:2013cia}
X.-N. Wang and Y.~Zhu,
\newblock Phys. Rev. Lett. {\bf 111}, 062301 (2013), 1302.5874.

\bibitem{Cao:2017qpx}
S.~Cao and A.~Majumder,
\newblock (2017), 1712.10055.

\bibitem{Cao:2017zih}
JETSCAPE, S.~Cao {\em et~al.},
\newblock Phys. Rev. {\bf C96}, 024909 (2017), 1705.00050.

\bibitem{Armesto:2009fj}
N.~Armesto, L.~Cunqueiro, and C.~A. Salgado,
\newblock Eur. Phys. J. {\bf C63}, 679 (2009), 0907.1014.

\bibitem{Eskola:2016oht}
K.~J. Eskola, P.~Paakkinen, H.~Paukkunen, and C.~A. Salgado,
\newblock Eur. Phys. J. {\bf C77}, 163 (2017), 1612.05741.

\bibitem{deFlorian:2011fp}
D.~de~Florian, R.~Sassot, P.~Zurita, and M.~Stratmann,
\newblock Phys. Rev. {\bf D85}, 074028 (2012), 1112.6324.

\bibitem{Kovarik:2015cma}
K.~Kovarik {\em et~al.},
\newblock Phys. Rev. {\bf D93}, 085037 (2016), 1509.00792.

\bibitem{Sassot:2009sh}
R.~Sassot, M.~Stratmann, and P.~Zurita,
\newblock Phys. Rev. {\bf D81}, 054001 (2010), 0912.1311.

\bibitem{deFlorian:2009vb}
D.~de~Florian, R.~Sassot, M.~Stratmann, and W.~Vogelsang,
\newblock Phys. Rev. {\bf D80}, 034030 (2009), 0904.3821.

\bibitem{dEnterria:2003xac}
D.~G. d'Enterria,
\newblock (2003), nucl-ex/0302016.

\bibitem{Cacciari:2008gp}
M.~Cacciari, G.~P. Salam, and G.~Soyez,
\newblock JHEP {\bf 04}, 063 (2008), 0802.1189.

\bibitem{Kang:2017mda}
Z.-B. Kang, F.~Ringer, and W.~J. Waalewijn,
\newblock JHEP {\bf 07}, 064 (2017), 1705.05375.

\bibitem{Banfi:2006hf}
A.~Banfi, G.~P. Salam, and G.~Zanderighi,
\newblock Eur. Phys. J. {\bf C47}, 113 (2006), hep-ph/0601139.

\bibitem{Cal:2019hjc}
P.~Cal, F.~Ringer, and W.~J. Waalewijn,
\newblock (2019), 1901.06389.

\bibitem{Acharya:2018uvf}
ALICE, S.~Acharya {\em et~al.},
\newblock JHEP {\bf 10}, 139 (2018), 1807.06854.

\bibitem{Aaboud:2018hpb}
ATLAS, M.~Aaboud {\em et~al.},
\newblock Phys. Rev. {\bf C98}, 024908 (2018), 1805.05424.

\bibitem{Sirunyan:2018qec}
CMS, A.~M. Sirunyan {\em et~al.},
\newblock Phys. Rev. Lett. {\bf 121}, 242301 (2018), 1801.04895.

\bibitem{Spousta:2015fca}
M.~Spousta and B.~Cole,
\newblock Eur. Phys. J. {\bf C76}, 50 (2016), 1504.05169.

\bibitem{Sirunyan:2017qhf}
CMS, A.~M. Sirunyan {\em et~al.},
\newblock Phys. Lett. {\bf B785}, 14 (2018), 1711.09738.

\bibitem{Aaboud:2018anc}
ATLAS, M.~Aaboud {\em et~al.},
\newblock Phys. Lett. {\bf B789}, 167 (2019), 1809.07280.

\bibitem{Adam:2015doa}
ALICE, J.~Adam {\em et~al.},
\newblock JHEP {\bf 09}, 170 (2015), 1506.03984.

\bibitem{Adamczyk:2017yhe}
STAR, L.~Adamczyk {\em et~al.},
\newblock Phys. Rev. {\bf C96}, 024905 (2017), 1702.01108.

\bibitem{Khachatryan:2016odn}
CMS, V.~Khachatryan {\em et~al.},
\newblock JHEP {\bf 04}, 039 (2017), 1611.01664.

\end{thebibliography}

\end{document}